\documentclass[10pt,conference]{IEEEtran}
\usepackage{amsfonts}
\usepackage{graphicx}
\usepackage{amsmath}
\usepackage{amssymb}
\usepackage{verbatim}
\newtheorem{thm}{Theorem}

\begin{document}
\title{Capacity Bounds of Half-Duplex Gaussian Cooperative Interference Channel}
\author{
\authorblockN{Yong Peng and Dinesh Rajan, \emph{Senior Member, IEEE}}
}

\maketitle \footnotetext[1]{The authors are with the Department of
Electrical Engineering at Southern Methodist University, Dallas,
Texas. Email: \{ypeng,rajand\}@engr.smu.edu. This work has been
supported in part by the National Science Foundation through grant
CCF 0546519.}
\begin{abstract}
In this paper, we investigate the half-duplex cooperative
communication scheme of a two user Gaussian interference channel. We
develop achievable region and outer bound for the case when the
system allow either transmitter or receiver cooperation. We show
that by using our transmitter cooperation scheme, there is
significant capacity improvement compare to the previous results
\cite{Shum,Fawaz}, especially when the cooperation link is strong.
Further, if the cooperation channel gain is infinity, both our
transmitter and receiver cooperation rates achieve their respective
outer bound. It is also shown that transmitter cooperation provides
larger achievable region than receiver cooperation under the same
channel and power conditions.
\end{abstract}
\begin{keywords}
Multi-user capacity, cooperative communications, relay channel,
dirty paper coding, Wyner-Ziv compress-and-forward.
\end{keywords}
\section{INTRODUCTION}
In wireless ad hoc networks, spatially dispersed radio terminals can
exploit cooperative diversity~\cite{Sendonaris,Laneman1} by relaying
signals for each other. With cooperation, different clusters of
terminals can act like transmit/receive antenna arrays and achieve increased
spatial diversity and throughput by joint encoding and/or decoding.

The capacity of the two-user Gaussian interference channel (IC) is
an open problem for  many years and is completely known only in some
special cases (\emph{e.g.}, in the strong interference
case~\cite{Sato}). The capacity region has been studied under
various cooperative strategies. Most of these schemes assume that
nodes operate in full-duplex mode. A coding scheme for transmitter
cooperation using decode-and-forward (DF) for relaying and dirty
paper coding (DPC) for codeword generation is proposed
in\cite{Madsen_tx}. Compress-and-forward~(CF) and DF relaying
strategies for receiver cooperation are proposed in~\cite{Madsen_rx}
and generalized to both transmitter and receiver
 cooperation in~\cite{Madsen}. A comparison of
 different coding schemes for transmitter cooperation in terms of
 the relative geometry of transmit and receive clusters is given in~\cite{Ng1}.
 The sum rate capacity with transmitter only, receiver only and both
 transmitter and receiver cooperation  is studied in~\cite{Ng2}. By using
 DF and DPC at the cooperative transmitters and Wyner-Ziv CF at the
 receivers and assuming equal power gain for all channels, the  proposed scheme in~\cite{Ng2}
 is shown to have significant capacity gain over strong IC~\cite{Sato}.
While full-duplex cooperative IC has been significantly studied, 
only limited results are known in the half-duplex scenario.
Cooperative
 diversity with transmitter cooperation for fading channels is
 considered in~\cite{Laneman1}. A 2-phase transmitter cooperation
 scheme using DF and the so called recycling DPC (RDPC) is
 introduced in~\cite{Shum}: Similar schemes are also proposed in~\cite{Fawaz},
 where the transmitters have additional flexibility in choosing the order of DPC.

In this paper, we compute bounds on the capacity of two user Gaussian
 IC in two different scenarios: i) transmitter cooperation (TC) and ii) receiver cooperation (RC).
Specifically, we allow all nodes to operate in half-duplex mode
only, which requires simpler and cheaper hardware.

In TC, the two transmit nodes serve as relays to each other. We
assume that the channel gain between the two transmitters  is much
higher than the others. In this case, it is well known that DF
strategy is superior~\cite{Kramer,Anders}. Thus, in this paper we
derive the achievable region with TC using only the DF strategy. We
show that the achievable region of the proposed TC strategy is
strictly larger than the results in~\cite{Shum,Fawaz}, especially
when the cooperation link is strong. In case when the cooperation
channel gain is infinity, the proposed achievable region achieves
the system upper bound. In contrast, for the schemes
in~\cite{Shum,Fawaz}, there is a large performance gap between the
lower and upper bounds.

In RC, the two receive nodes serve as relays to each other. In this
case, we assume that the relay to destination channel is strong for
RC, and CF~\cite{Kramer} is preferable at the relays. Thus, to
derive the achievable region with RC, we only consider the CF
strategy. The proposed scheme achieves the corresponding MIMO
multiple access channel (MAC) capacity \cite{Telatar} when the
cooperation channel gain is infinity. To the best of our knowledge,
the achievable rate with RC has not been studied under the
half-duplex assumption. We also show that under identical channel
conditions and equal transmit power constraints on all nodes, TC
achieves larger rates than RC.

\section{SYSTEM MODEL} \label{sec:models}
\begin{figure}[htp]
\begin{center}
\includegraphics[width=3.4in]{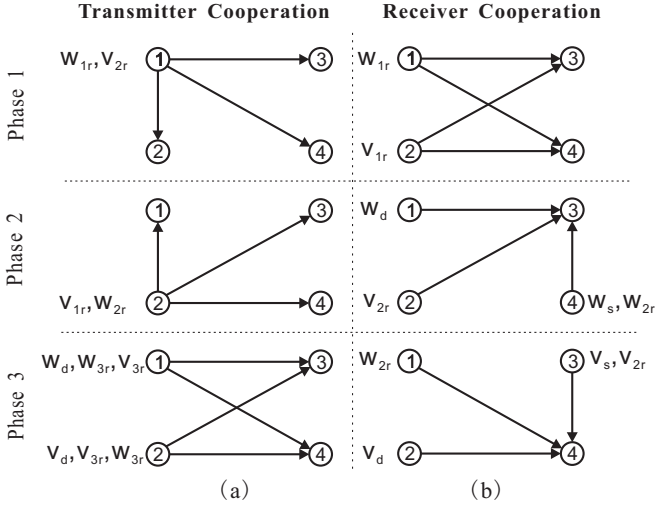}
\caption{System models of half-duplex Gaussian cooperative
interference channel with (a) transmitter cooperation and (b)
receiver cooperation.}\label{fig:model}
\end{center}
\end{figure}
Consider a two-transmitter two-receiver network shown in Fig.
\ref{fig:model}, where node $3$ is the intended receiver of node $1$
and node $4$ is the intended receiver of node $2$. The independent
messages transmitted by node $i,~i\in\{1,2,3,4\}$ are encoded into
$N$ complex symbols $x_i[1],x_i[2],\ldots,x_i[N]$, under the power
constraint $\frac{1}{N}\sum_{n=1}^{N}x_i[n]^2\leq P_i$. If the
messages transmitted by node $1$ and $2$ has a total alphabet of
$M_1$ and $M_2$ respectively, their respective rates are then
$R_1=\log M_1/N$ and $R_2=\log M_2/N$ bits/transmission. The channel
gain from node $i$ to node $k$ and $k>i$, is represented by a
complex constant $h_{ik}=c_{ik}e^{j\theta_{ik}}$. It is assumed that
all nodes have perfect knowledge of the channel gain and all the
phase offsets can be perfectly synchronized. Let $\mathbf{z}_i$
denote the \emph{i.i.d.} complex circularly symmetric Gaussian noise
process at node $i$, with the $n^{th}$ element
$z_i[n]\sim\mathbb{C}\mathcal{N}(0,1)$. We assume that the
communication is in a half-duplex fashion, \emph{i.e.}, each of the
nodes can be either in the transmit mode or the receive mode. For
TC, only the two transmit nodes (node 1 and 2) can cooperate with
each other while for RC, only the two receive nodes (node 3 and 4)
can cooperate with each other. It is also assumed that the
cooperation nodes are close together, \emph{i.e.}, $c_{12}$ and
$c_{34}$ are large compared to the other~$c_{ik}$'s. Further, we
define the following non-negative parameters satisfying
$\alpha_1+\alpha_2=1$, $\beta_1+\beta_2=1$, $\kappa_1+\kappa_2=1$,
$\gamma_1+\gamma_2=1$, $\mu_1+\mu_2+\mu_3=1$,
$\eta_1+\eta_2+\eta_3=1$ and $\lambda_1+\lambda_2+\lambda_3=1$. Also
define $\mathbf{g}_1=[c_{13}~c_{23}]$, $\mathbf{g}_2=[c_{14}~
c_{24}]$, $\mathbf{h}_1=[c_{13}~ c_{14}]$ and $\mathbf{h}_2=[c_{23}~
c_{24}]$. Let $C(x)=\log(1+x)$.

\section{TRANSMITTER COOPERATION}

\subsection{Achievable Rates}
\begin{thm} For the half-duplex Gaussian interference
channel where the transmitters can cooperate with each other, all
rate pairs $(R_1^{T_X},R_2^{T_X})$ satisfying
\begin{eqnarray}
R_1^{T_X}&\leq&\min\left\{R_{1,d}^{T_X}+R_{1,r_1}^{T_X},R_{1,d}^{T_X}+R_{1,r_2}^{T_X}\right\}\\
R_2^{T_X}&\leq&\min\left\{R_{2,d}^{T_X}+R_{2,r_1}^{T_X},R_{2,d}^{T_X}+R_{2,r_2}^{T_X}\right\}
\end{eqnarray}
are achievable, where $R_{i,d}^{T_X}$ is given by (\ref{eq:t_R1d})
and (\ref{eq:t_R2d}), $R_{i,r_1}^{T_X}$ is given by
(\ref{eq:t_R1r1}) and (\ref{eq:t_R2r1}), and $R_{i,r_2}^{T_X}$ is
given by (\ref{eq:t_R1r2}) and (\ref{eq:t_R2r2}).
\end{thm}

{\em Proof:} We construct a 3-phase transmission strategy as shown
in Fig. \ref{fig:model}-(a), to show the achievability. Let $w_i$'s
and $v_i$'s be the messages intended to node 3 and 4 respectively.
The specific message sent in each
phase is detailed in Fig. \ref{fig:model}-(a). 
 In phase 1 and 2, the two source nodes transmit messages
$w_{1r}$ and $v_{1r}$ to each other, and $w_{2r}$ and $v_{2r}$ to
the receive nodes by broadcasting their signals using DPC. In phase
3, after the sources exchanged their information, the system is
equivalent to a two user 2-transmit-1-receive antenna MIMO BC. The
source nodes can then jointly broadcast $w_{3r}$ and $v_{3r}$ to the
receivers using DPC \cite{Shamai}. Further, the two source nodes can
also send $w_d$ and $v_d$ in phase 3, respectively. Due to the
limited space, we only outline the results at each phase.

{\underline{Transmission Scheme:}} The transmission is divided into
3 phases as shown in Fig.~\ref{fig:model}-(a), with time
portion $\lambda_1$ $\lambda_2$ and $\lambda_3$.
~In \emph{Phase 1}, node $1$ is in transmit mode and all other nodes
are in receive mode. The received signal at node~$i$
$y_i[n]=h_{1i}x_1[n]+z_i[n],~ n\in
\{1,2,\ldots,$$\lfloor\lambda_1N\rfloor\}, i =2,3,$ and~4. In
\emph{Phase 2}, node $2$ is in the transmit mode and all the other
nodes are in receive mode.  In \emph{Phase 3}, nodes~$1$ and~$2$ are
in transmit mode and nodes~$3$ and~$4$ are in the receive mode. The
received signal in phases~2 and~3 can easily be expressed similar to
phase~1.

{\underline{Outline of Achievability:}}

1) \emph{Phase 1:} If $c_{13}>c_{14}$, generate codeword
$\mathbf{X}_1(v_{2r})$ with length $\lambda_1N,~N\rightarrow \infty$
and power $\alpha_2P_1^{(1)},~P_1^{(1)}=\kappa_1P_1/\lambda_1$.
Given~$\mathbf{X}_1(v_{2r})$, use DPC to generate
$\mathbf{X}_1(w_{1r})$ with length $\lambda_1N$ and power
$\alpha_1P_1^{(1)}$. Otherwise, do DPC with the reverse order. Since
$v_{2r}$ is known to node 2, it can subtract
$\mathbf{X}_1(v_{2r})$ and decode $w_{1r}$ if the rate of $w_{1r}$
satisfies~\cite{Shum}
\begin{equation}\label{eq:t_R1r1}
R_{1,r_1}^{T_X}\leq\lambda_1C\left({c_{12}^2\alpha_1P_1^{(1)}}\right).
\end{equation}
Node 3 can decode $v_{2r}$ if the rate of $v_{2r}$ satisfies
\begin{equation}\label{eq:t_R21}
R_{2,1}^{T_X}\leq\left\{\begin{aligned}&\lambda_1C\left({c_{14}^2\alpha_2P_1^{(1)}}/
{(1+c_{14}^2\alpha_1P_1^{(1)})}\right),&&\text{if}~c_{13}> c_{14}\\
&\lambda_1C\left({c_{14}^2\alpha_2P_1^{(1)}}\right),&&\text{otherwise}
\end{aligned}.\right.\nonumber
\end{equation}

 2) \emph{Phase 2:} If $c_{24}>c_{23}$, generate codeword
$\mathbf{X}_1(w_{2r})$ with length $\lambda_2N$ and power
$\beta_2P_2^{(1)},~P_2^{(1)}=\gamma_1P_2/\lambda_2$. Given
$\mathbf{X}_1(w_{2r})$, use DPC to generate $\mathbf{X}_2(v_{1r})$
with length $\lambda_2N$ and power $\beta_1P_2^{(1)}$. Otherwise, do
DPC in the reverse order. Node 1 can decode $v_{1r}$ if the rate of
$v_{1r}$ satisfies \cite{Shum}
\begin{equation}\label{eq:t_R2r1}
R_{2,r_1}^{T_X}\leq\lambda_2C\left({c_{12}^2\beta_1P_2^{(1)}}\right)
\end{equation}
and node 3 can decode $w_{2r}$ if the rate of $w_{2r}$ satisfies
\begin{equation}\label{eq:t_R12}
R_{1,2}^{T_X}\leq\left\{\begin{aligned}&\lambda_2C\left({c_{23}^2\beta_2P_2^{(1)}}/
{(1+c_{23}^2\beta_1P_2^{(1)})}\right),&&\text{if}~c_{24}> c_{23}\\
&\lambda_2C\left({c_{23}^2\beta_2P_2^{(1)}}\right),&&\text{otherwise}
\end{aligned}\right..\nonumber
\end{equation}

 3) \emph{Phase 3:} After phase 1 and 2, $v_{1r}$ and $w_{1r}$ have been exchanged between
 the sources. Node 1 and 2 can then sent messages jointly using the coding scheme of a two user 2-transmit-1-receive antenna MIMO
 BC \cite{Shamai}. The problem now is to find the optimal covariance matrices
 of the two transmit signals for both receive node 3 and 4. In
 \cite{Goldsmith}, a simple method of generating MIMO BC covariance
 matrices is proposed by transforming the covariance matrices from its dual, MIMO
 MAC. We use this method to find the covariance matrices $\Sigma_i$ and $\Sigma_i'$ in our coding scheme.

If $c_{13}+c_{23}>c_{14}+c_{24}$, generate codeword
$\mathbf{X}_2(v_d)$ with length $\lambda_3N$ and power
$\eta_1P_2^{(2)},~P_2^{(2)}=\gamma_2P_2/\lambda_3$ at node 2.
Generate codeword $\mathbf{X}_1{(v_{3r})}$ and
$\mathbf{X}_2{(v_{3r})}$ with length $\lambda_3N$ at node 1 and 2
respectively with covariance matrix $\Sigma_2$, where $\Sigma_2$ can
be found by using the results given in \cite{Goldsmith}. Let
$\mathbf{B}_1=\mathbf{I}+\mathbf{h}_2^T\mathbf{h}_2(\mu_3P_1^{(2)}+\eta_2P_2^{(2)}),~P_1^{(2)}=\kappa_2P_1/\lambda_3$,
then $\Sigma_1=\mathbf{B}_1^{-1}(\mu_2P_1^{(2)}+\eta_3P_2^{(2)})$.
Let $A_2=1+\mathbf{h}_2\Sigma_1\mathbf{h}_2^T$, then
$\Sigma_2=A_2(\mu_3P_1^{(2)}+\eta_2P_2^{(2)})\mathbf{I}$. Given
$\mathbf{X}_2(v_d)$ and $\mathbf{X}_1{(v_{3r})}$, use DPC to
generate codeword $\mathbf{X}_1(w_d)$ with length $\lambda_3N$ and
power $\mu_1P_1^{(2)}$ at node~1. Generate codeword
$\mathbf{X}_1{(w_{3r})}$ and $\mathbf{X}_2(w_{3r})$ with length
$\lambda_3N$ at node 1 and 2 respectively with covariance matrix
$\Sigma_1$. If $c_{13}+c_{23}\leq c_{14}+c_{24}$, do DPC with the
reverse order. Note that in this case, the covariance matrix becomes
$\Sigma_1'={\mathbf{B}}_1'^{-1}(\mu_3P_1^{(2)}+\eta_2P_2^{(2)})$,
where
${\mathbf{B}}_1'=\mathbf{I}+\mathbf{h}_1^T\mathbf{h}_1(\mu_2P_1^{(2)}+\eta_3P_2^{(2)})$
and $\Sigma_2'={A}_2'(\mu_2P_1^{(2)}+\eta_3P_2^{(2)})\mathbf{I}$,
where ${A}_2'=1+\mathbf{h}_1\Sigma_1\mathbf{h}_1^T$\footnote{It is
easy to show that if we let
$\Sigma_1=\Sigma_2'=\text{diag}\left\{\mu_2P_1^{(2)},\eta_3P_2^{(2)}\right\}$
and
$\Sigma_1'=\Sigma_2=\text{diag}\left\{\mu_3P_1^{(2)},\eta_2P_2^{(2)}\right\}$,
the achievable rates of our scheme reduces to the rates
given by parallel coding DPC in \cite{Fawaz} (or RDPC in~\cite{Shum}, if we further restrict the condition to $c_{13}\leq
c_{14}$ and $c_{24}\leq c_{23}$). Note that using the above
covariance matrices pairs is equivalent to the case assuming random
phase shifts for different channels, \emph{i.e.}, the received
signal from different transmitters can not be synchronized.}. Node 3
first decodes $w_{3r}$, it can do so if the rate of $w_{3r}$
satisfies
\begin{equation}\label{eq:t_R13}
R_{1,3}^{T_X}\leq\left\{\begin{aligned}&\lambda_3C\left(\frac{\mathbf{g}_1\Sigma_1\mathbf{g}_1^{T}}{c_{13}^2\mu_1P_1^{(2)}}\right),\quad~
\text{if}~c_{13}+ c_{23}>c_{14}+c_{24}\\
&\lambda_3C\left(\frac{\mathbf{g}_1\Sigma_2'\mathbf{g}_1^T}{1+\mathbf{g}_1\Sigma_1'\mathbf{g}_1^T+c_{13}^2\mu_1P_1^{(2)}+c_{23}^2\eta_1P_2^{(2)}}\right)
,\\&\quad\quad\quad\quad\quad\quad\quad\quad\quad\quad\quad\quad\quad\quad~~\quad\text{otherwise}
\end{aligned}\right..\nonumber
\end{equation}
Node 3 then decodes $w_d$, if the rate of $w_d$ satisfies
\begin{equation}\label{eq:t_R1d}
R_{1,d}^{T_X}\leq\left\{\begin{aligned}&\lambda_3C\left({c_{13}^2\mu_1P_1^{(2)}}\right),\quad\text{if}~c_{13}+ c_{23}>c_{14}+c_{24}\\
&\lambda_3C\left(\frac{c_{13}^2\mu_1P_1^{(2)}}{1+\mathbf{g}_1\Sigma_1'\mathbf{g}_1^T+c_{23}^2\eta_1P_2^{(2)}}\right),\text{otherwise}
\end{aligned}\right..
\end{equation}
 After decoding $w_{2r}$ and $w_{3r}$, node 3 can finally decode
$w_{1r}$ if the rate of $w_{1r}$ satisfies
\begin{equation}\label{eq:t_R1r2}
R_{1,r_2}^{T_X}\leq R_{1,1}^{T_X}+R_{1,2}^{T_X}+R_{1,3}^{T_X}
\end{equation}
where
\begin{equation}
R_{1,1}^{T_X}\leq\left\{\begin{aligned}&\lambda_1C\left({c_{13}^2\alpha_1P_1^{(1)}}\right),&&\text{if}~c_{13}> c_{14}\\
&\lambda_1C\left(\frac{c_{13}^2\alpha_1P_1^{(1)}}{1+c_{13}^2\alpha_2P_1^{(1)}}\right),&&\text{otherwise}
\end{aligned}\right..\nonumber
\end{equation}
Similarly, node 4 first decodes $v_{3r}$, if the rate of $v_{3r}$
satisfies
\begin{equation}
R_{2,3}^{T_X}\leq\left\{\begin{aligned}
&\lambda_3C\left(\frac{\mathbf{g}_2\Sigma_2\mathbf{g}_2^T}{1+\mathbf{g}_2\Sigma_1\mathbf{g}_2^T+c_{14}^2\mu_1P_1^{(2)}+c_{24}^2\eta_1P_2^{(2)}}\right),
\\&\quad\quad\quad\quad\quad\quad\quad\quad\quad\quad~~\text{if}~c_{13}+ c_{23}>c_{14}+c_{24}\\
&\lambda_3C\left({\mathbf{g}_2\Sigma_1'\mathbf{g}_2^T}/{c_{24}^2\eta_1P_2^{(2)}}\right),\quad\text{otherwise}
\end{aligned}\right..\nonumber
\end{equation}
Node 4 can then decode $v_d$ if the rate of $v_d$ satisfies
\begin{equation}\label{eq:t_R2d}
R_{2,d}^{T_X}\leq\left\{\begin{aligned}&\lambda_3C\left(\frac{c_{24}^2\eta_1P_2^{(2)}}
{1+\mathbf{g}_2\Sigma_1\mathbf{g}_2^T+c_{14}^2\mu_1P_1^{(2)}}\right),\\&~~\quad\quad\quad\quad\quad\quad\quad\quad\quad\quad\text{if}~c_{13}+
c_{23}>c_{14}+c_{24}\\
&\lambda_3C\left({c_{24}^2\eta_1P_2^{(2)}}\right),\quad\quad~~\thinspace\text{otherwise}
\end{aligned}\right..
\end{equation}
 After decoding $v_{2r}$ and $v_{3r}$, node 4 can decode
$v_{1r}$ if the rate of $v_{1r}$ satisfies\footnote{Note that for
the transmission order given in
Fig. \ref{fig:model}-(a), $v_{1r}$ is encoded and transmitted in
phase 2, the receiver can decode it only after $v_{2r}$ and $v_{3r}$
been decoded at phase 1 and 3 of the next transmission block.}
\begin{equation}\label{eq:t_R2r2}
R_{2,r_2}^{T_X}\leq R_{2,1}^{T_X}+R_{2,2}^{T_X}+R_{2,3}^{T_X}
\end{equation}
where
\begin{equation}
R_{2,2}^{T_X}\leq\left\{\begin{aligned}&\lambda_2C\left({c_{24}^2\beta_1P_2^{(1)}}\right),&&\text{if}~c_{24}> c_{23}\\
&\lambda_2C\left(\frac{c_{24}^2\beta_1P_2^{(1)}}{1+c_{24}^2\beta_2P_2^{(1)}}\right),&&\text{otherwise}
\end{aligned}\right..\nonumber
\end{equation}

\subsection{Outer Bound}
For TC, when $c_{12}\rightarrow \infty$, the system becomes a two
user 2-transmit-1-receive antenna MIMO BC. The capacity region of
this MIMO BC~\cite{Shamai} is an outer bound on achievable rate.
Further, when one user is silent, the achievable rate for the active
user is bounded by the single user half-duplex relay channel
max-flow-min-cut bound~\cite{Anders}. Hence, with TC, the set of
achievable rate pairs $(R_1^+,R_2^{+})$ satisfies
\begin{align}
R_i^+&\leq\underset{{0\leq \rho_i \leq 1}
}\max\min\{R_{i,1}^+(\rho_i),~R_{i,2}^+(\rho_i)\},~i=1,2\label{eq:single1}\\
R_1^++R_2^+&\leq \underset{\forall P_1+P_2<P} \bigcup
\mathbf{C}(\mathbf{g}_1^TP_1\mathbf{g}_1+\mathbf{g}_2^TP_2\mathbf{g}_2).
\label{eq:MIMO_BC}
\end{align}
where $\mathbf{C}(\mathbf{x})=\log |\mathbf{I}+\mathbf{x}|$ and
$\bigcup$ is the union of all rates with any power allocations $P_1$
and $P_2$ that satisfies the total power constraint $P$, and
\begin{align}
R_{1,1}^+(\rho_1)=&\alpha_1C\left(c_{12}^2+c_{13}^2P_1\right)+\alpha_2C\left((1-\rho_1)c_{13}^2P_1\right)\nonumber\\
R_{1,2}^+(\rho_1)=&\alpha_1C\left(c_{13}^2P_1\right)+ \alpha_2C\left(c_{13}^2P_1+c_{23}^2P_2+2\varphi_1~\right)\nonumber\\
R_{2,1}^+(\rho_2)=&\alpha_1C\left(c_{12}^2+c_{24}^2P_2\right)+\alpha_2C\left((1-\rho_2)c_{24}^2P_2\right)\nonumber\\
R_{2,2}^+(\rho_2)=&\alpha_1C\left(c_{24}^2P_2\right)+
\alpha_2C\left(c_{14}^2P_1+c_{24}^2P_2+2\varphi_2~\right)\nonumber
\end{align}
where $\varphi_1=\sqrt{\rho_1c_{13}^2c_{23}^2P_1P_2}$ and
$\varphi_2=\sqrt{\rho_2c_{14}^2c_{24}^2P_1P_2}$,

\section{RECEIVER COOPERATION}
\subsection{Achievable Rates}
\begin{thm}
For the half-duplex Gaussian interference channel where the
receivers can cooperate with each other, all rate pairs
$(R_1^{Rx},R_2^{Rx})$ satisfying
\begin{eqnarray}
R_1^{R_X}&\leq& R_{1,d}^{R_X}+R_{1,r_1}^{R_X}+R_{1,r_2}^{R_X}\\
R_2^{R_X}&\leq& R_{2,d}^{R_X}+R_{2,r_1}^{R_X}+R_{2,r_2}^{R_X}
\end{eqnarray}
are achievable, where $R_{i,d}^{R_X}$ is given by (\ref{eq:r_R1d})
and (\ref{eq:r_R2d}), $R_{i,r_1}^{R_X}$ is given by the inequalities
from (\ref{eq:r_R1r1}) to (\ref{eq:r_R2r1}), and $R_{i,r_2}^{R_X}$
is given by (\ref{eq:r_R1r2}) and (\ref{eq:r_R2r2}).
\end{thm}

\emph{Proof:} The 3-phase RC scheme is shown in Fig.
\ref{fig:model}-(b). In phase 1, the signals from node 1 and 2 are
received at node 3 and 4. Rather than decoding the signals, the two
receive nodes exchange information in phase 2 and 3 by sending each
other a compressed version of what they received. The receive nodes
then perform decoding by using the aggregation of the compressed
signal and the signal directly received in phase 1. Let $w_i$'s and
$v_i$'s
be the messages intended to node 3 and 4 respectively. The specific message sent in each phase is detailed in Fig. \ref{fig:model}-(b).  
We outline the coding scheme as follows.

{\underline {Transmission Scheme:}} In \emph{Phase 1}, nodes~$3$
and~$4$ are in receive mode and nodes~$1$ and $2$ are in transmit
mode. Again, since the expressions of the received signals can be
easily shown, we omit them due to limited space. In \emph{Phase 2},
node $3$ is in receive mode and all the other nodes are in
transmit mode. 
In \emph{Phase 3}, node~$4$ is in receive mode and all the other
nodes are in
transmit mode. 

{\underline{Outline of Achievability:}}

\emph{Phase 1:} At nodes~1 and~2, generate $\lambda_1N$ length codewords
$\mathbf{X}_1(w_{1r})$ and $\mathbf{X}_2({v_{1r}})$ with powers
$P_1^{(1)}=\mu_1P_1\lambda_1$ and $P_2^{(1)}=\eta_1P_2\lambda_1$ respectively.

\emph{Phase 2:} At node~1 and~2, generate $\lambda_2N$ length
codewords $\mathbf{X}_1({w_{d}})$ and $\mathbf{X}_2({v_{2r}})$ with
powers $P_1^{(2)}=\mu_2P_1/\lambda_2$ and
$P_2^{(2)}=\eta_2P_2/\lambda_2$ respectively. At node~$4$, generate
$\lambda_2N$ length codewords $\mathbf{X}_2(w_s)$ and
$\mathbf{X}_2(w_{2r})$\footnote{Note that $w_{2r}$ is the message
decoded at phase 3 of the previous block. It is re-encoded as
$\mathbf{X}_2(w_{2r})$ and relayed to the intended receiver.} with
power $P_4^{(1)}=\alpha_1P_4/\lambda_2$ and
$P_4^{(2)}=\alpha_2P_4/\lambda_2$ respectively. Node 3 first decode
$w_{2r}$, if the rate of $w_{2r}$ satisfies
\begin{equation}\label{eq:R12r2}
R_{1,2r_2}^{R_X}\leq
\lambda_2C\left(\frac{c_{34}^2P_4^{(2)}}{1+c_{13}^2P_1^{(2)}+c_{23}^2P_2^{(2)}+c_{34}^2P_4^{(1)}}\right).
\end{equation}
Node 3 can then decode~$w_s$, if the rate of~$w_s$ satisfies
\begin{equation}
R_{1,s}^{R_X}\leq \lambda_2C\left({c_{34}^2P_4^{(1)}}/
{(1+c_{13}^2P_1^{(2)}+c_{23}^2P_2^{(2)})}\right)
\end{equation}
and decode $v_{2r}$ and $w_d$, if their respective rates satisfy
\begin{align}
R_{2,2r_1}^{R_X}&\leq
\lambda_2C\left({c_{23}^2P_2^{(2)}}/({1+c_{13}^2P_1^{(2)}})\right)\label{eq:R22r1}\\
R_{1,d}^{R_X}&\leq\lambda_2C\left(c_{13}^2P_1^{(2)}\right).\label{eq:r_R1d}
\end{align}

\emph{Phase 3:} At nodes~1 and~2, generate $\lambda_3N$ length
codewords $\mathbf{X}_1({w_{2r}})$ and $\mathbf{X}_2({v_{d}})$ with
powers $P_1^{(3)}=\mu_3P_1/\lambda_3$ and
$P_2^{(3)}=\eta_3P_2/\lambda_3$ respectively. At node~3, generate
$\lambda_3N$ length codewords $\mathbf{X}_1(v_s)$ and
$\mathbf{X}_2(v_{2r})$ with powers $P_3^{(1)}=\beta_1P_3/\lambda_3$
and $P_3^{(2)}=\beta_2P_3/\lambda_3$ respectively. Node 4 can decode
$v_{2r}$ if
\begin{equation}\label{eq:R22r2}
R_{2,2r_2}^{R_X}\leq
\lambda_3C\left(\frac{c_{34}^2P_3^{(2)}}{1+c_{14}^2P_1^{(3)}+c_{24}^2P_2^{(3)}+c_{34}^2P_3^{(1)}}\right).
\end{equation}
Combining (\ref{eq:R22r1}) and (\ref{eq:R22r2}), node 4 can decode
$v_{2r}$ if
\begin{equation}\label{eq:r_R2r2}
R_{2,r_2}^{R_X}\leq\min\left\{\max\left(R_{2,2r_1}^{R_X}\right),\max\left(R_{2,2r_2}^{R_X}\right)\right\}.
\end{equation}
Node 4 can then decodes $v_{s}$, if the rate of $v_s$ satisfies
\begin{equation}
R_{2,s}^{R_X}\leq \lambda_3C\left({c_{34}^2P_3^{(1)}}/
{(1+c_{14}^2P_1^{(3)}+c_{24}^2P_2^{(3)})}\right).
\end{equation}
After decoding $v_{2r}$ and $v_s$, node 4 decodes $w_{2r}$ if
\begin{equation}\label{eq:R12r1}
R_{1,2r_1}^{R_X}\leq
\lambda_3C\left({c_{14}^2P_1^{(3)}}/{(1+c_{24}^2P_2^{(3)})}\right).
\end{equation}
Combining (\ref{eq:R12r1}) and (\ref{eq:R12r2}), node 3 can decode
$w_{2r}$ if
\begin{equation}\label{eq:r_R1r2}
R_{1,r_2}^{R_X}\leq\min\left\{\max\left(R_{1,2r_1}^{R_X}\right),\max\left(R_{1,2r_2}^{R_X}\right)\right\}.
\end{equation}
Finally, node 4 can decode $v_d$ if the rate of $v_d$ satisfies
\begin{equation}\label{eq:r_R2d}
R_{2,d}^{R_X}\leq\lambda_3C\left(c_{24}^2P_2^{(3)}\right).
\end{equation}

We now consider the decoding of $w_{1r}$ and $v_{1r}$. By decoding
$w_s$ and $v_s$, a compressed version of the signals received in
phase 1 have been exchanged between the receivers. Let $\sigma_1^2$
and $\sigma_2^2$ be the compression noise of the received signal at
node 3 and 4 respectively. Using similar derivations as in
\cite{Ng2}, $\sigma_1^2$ and $\sigma_2^2$ are given by
\begin{align}
\sigma_1^2&=\frac{\left(1+\mathbf{g}_1\Sigma_x\mathbf{g}_1^{T}\right)\left(1+\mathbf{g}_2\Sigma_x\mathbf{g}_2^{T}\right)
-\left(\mathbf{g}_1\Sigma_x\mathbf{g}_2^{T}\right)^2}{\left(2^{R_{2,s}^{R_X}/{\lambda_1}}-1\right)\left(1+\mathbf{g}_2\Sigma_x\mathbf{g}_2^{T}\right)}\\
\sigma_2^2&=\frac{\left(1+\mathbf{g}_2\Sigma_x\mathbf{g}_2^{T}\right)\left(1+\mathbf{g}_1\Sigma_x\mathbf{g}_1^{T}\right)
-\left(\mathbf{g}_1\Sigma_x\mathbf{g}_2^{T}\right)^2}{\left(2^{R_{1,s}^{R_X}/{\lambda_1}}-1\right)\left(1+\mathbf{g}_1\Sigma_x\mathbf{g}_1^{T}\right)}
\end{align}
where $\Sigma_x=\text{diag}\left\{P_1^{(1)},~P_2^{(1)}\right\}$ is a
${2\times 2}$ diagonal matrix.

 As discussed in \cite{Ng2}, since each receiver has a noisy
version of the received signal of the other receiver, the network is
equivalent to an IC with two receive antennas at each receiver.
After normalizing the noise power to 1 for all receive ``antennas",
the equivalent channel gains between the transmit and receive node
pairs are given as
$\mathbf{c}_{13}=[c_{13}~\sqrt{\zeta_2}c_{14}]^T$,
$\mathbf{c}_{23}=[c_{23}~\sqrt{\zeta_2}c_{24}]^T$,
$\mathbf{c}_{14}=[\sqrt{\zeta_1}c_{13}~c_{14}]^T$ and
$\mathbf{c}_{24}=[\sqrt{\zeta_1}c_{23}~c_{24}]^T$, where
${\zeta}_i=1/(1+\sigma_i^2)$. Let $
\mathbf{{SNR}}_1=\mathbf{c}_{13}\mathbf{c}_{13}^TP_1^{(1)}$, $
\mathbf{{INR}}_1=\mathbf{c}_{23}\mathbf{c}_{23}^TP_2^{(1)}$, $
\mathbf{{SNR}}_2=\mathbf{c}_{24}\mathbf{c}_{24}^TP_1^{(1)}$ and $
\mathbf{{INR}}_2=\mathbf{c}_{14}\mathbf{c}_{14}^TP_2^{(1)}. $

The capacity region of a 1-transmit-2-receive antennas IC is not
known except for the strong interference case \cite{Jafar}
($||\mathbf{c}_{14}||^2\geq ||\mathbf{c}_{13}||^2$ and
$||\mathbf{c}_{23}||^2\geq ||\mathbf{c}_{24}||^2$). In this case,
the messages $w_{1r}$ and $v_{1r}$ can be decoded if their
respective rate $R_{1,r1}^{R_X}$ and $R_{2,r1}^{R_X}$ satisfies
\cite{Jafar}
\begin{align}
R_{1,r1}^{R_X}&\leq \lambda_1\mathbf{C}\left(\mathbf{{SNR}}_1\right)\label{eq:r_R1r1}\\
R_{2,r1}^{R_X}&\leq\lambda_1\mathbf{C}\left(\mathbf{{SNR}}_2\right)\\
R_{1,r1}^{R_X}+R_{2,r1}^{R_X}&\leq \lambda_1 \underset{i=1,2}\min
\left\{\mathbf{C}\left(\mathbf{{SNR}}_i+\mathbf{{INR}}_i\right)\right\}.
\end{align}

When $||\mathbf{c}_{14}||^2\geq
||\mathbf{c}_{13}||^2,~||\mathbf{c}_{23}||^2<
||\mathbf{c}_{24}||^2$, node 4 can eliminate the interference by
completely decoding the message transmitted from node 1 and node 3
can decode by treating its interference as noise. Thus, the
achievable rates of $w_{1r}$ and $v_{1r}$ are respectively
\begin{align}
R_{1,r1}^{R_X}&\leq \lambda_1\log\left({\left|\mathbf{I}+\mathbf{SNR}_1+\mathbf{INR}_1\right|}/{\left|\mathbf{I}+\mathbf{INR}_1\right|} \label{eq:R1}\right)\\
R_{2,r1}^{R_X}&\leq \lambda_1\mathbf{C}\left(\mathbf{SNR}_2\right).
\end{align}
Similarly, when $||\mathbf{c}_{14}||^2<
||\mathbf{c}_{13}||^2,~||\mathbf{c}_{23}||^2\geq
||\mathbf{c}_{24}||^2$, the achievable rates of $w_{1r}$ and
$v_{1r}$ are respectively given by
\begin{align}
R_{1,r1}^{R_X}&\leq \lambda_1\mathbf{C}\left(\mathbf{SNR}_1\right)\\
R_{2,r1}^{R_X}&\leq\lambda_1\log\left({\left|\mathbf{I}+\mathbf{SNR}_2+\mathbf{INR}_2\right|}/{\left|\mathbf{I}+\mathbf{INR}_2\right|}\right)
\label{eq:r_R2r1}.
\end{align}

When $||\mathbf{c}_{14}||^2< ||\mathbf{c}_{13}||^2$ and
$||\mathbf{c}_{23}||^2< ||\mathbf{c}_{24}||^2$, node 3 and node 4
can decode $w_{1r}$ and $v_{1r}$ respectively by treating
interference as noise. The achievable rates of $w_{1r}$ and $v_{1r}$
are then given by~(\ref{eq:R1}) and~(\ref{eq:r_R2r1}).

\subsection{Outer Bound}
The single user upper bounds in~(\ref{eq:single1}) are also upper
bounds under RC. Further, if we let $c_{34}= \infty$, the channel
becomes a two user 1-transmit-2-receive antenna MIMO MAC. Thus, the
achievable region is also bounded by this MIMO MAC capacity, which
is given by \cite{Telatar}
\begin{equation}\label{eq:Rx_out}
R_1^++R_2^+\leq
\mathbf{C}(\mathbf{h}_1^TP_1\mathbf{h}_1+\mathbf{h}_2^TP_2\mathbf{h}_2).
\end{equation}

\section{NUMERICAL EXAMPLES}
We compare our achievable region to some known results through
numerical examples. We focus on the symmetric channel case (similar
results can be shown for the asymmetric case). We set the direct
channel gains as $c_{13}=c_{24}=1$, the cross channel gains as
$c_{14}=c_{23}=\sqrt{2}$ and the average power constraints
$P_i=5,~i=1,2,3,4$.

\begin{figure}[htp]
\begin{center}
\includegraphics[width=3.2in]{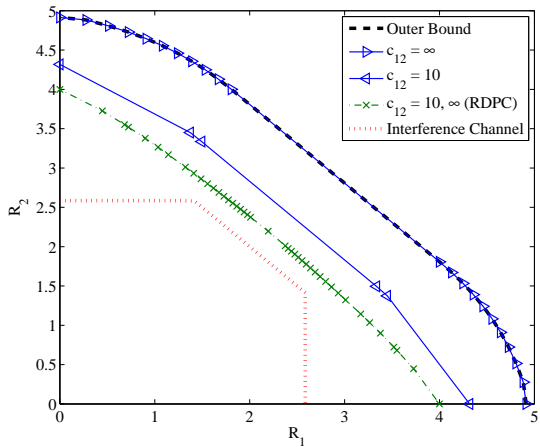}
\caption{Achievable regions for transmitter cooperation and RDPC.}
\label{fig:tx}
\end{center}
\end{figure}
Fig. \ref{fig:tx} compares the achievable region from our TC scheme
with RDPC \cite{Shum,Fawaz}. It is shown that the achievable region
using our TC scheme is significantly larger than using RDPC.
Further, the capacity gain of our TC scheme increases with the
cooperation channel gain: As we increase the cooperation channel
gain from $c_{12}=10$ to $\infty$, the achievable region meets the
outer bound. On the other hand, the achievable region of RDPC does
not increase as long as the cooperation channel is not a capacity
threshold (see equations (8) and (9) in \cite{Shum}). The achievable
regions are also compared to the capacity of a standard strong IC
(without node cooperation). It is clear that by allowing node
cooperation, the achievable region increases significantly.

\begin{figure}[htp]
\begin{center}
\includegraphics[width=3.2in]{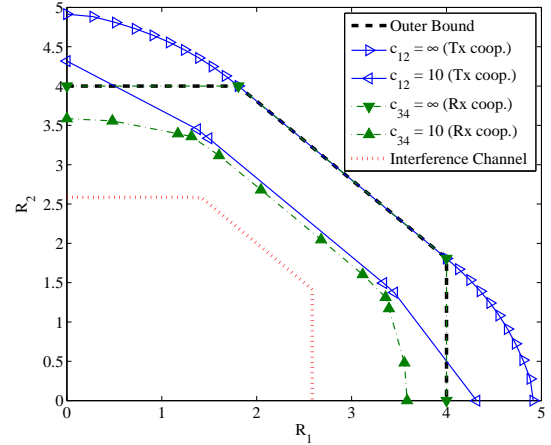}
\caption{Achievable regions for transmitter
 and receiver cooperation.}
\label{fig:rx}
\end{center}
\end{figure}
Fig. \ref{fig:rx} shows the achievable regions for both TC and RC.
Similar to TC, the achievable region of RC also increases with
cooperation channel gain. When $c_{34}=\infty$, the achievable
region of RC overlaps with the outer bound. The RC achievable region
is also compared with TC. When $c_{12}=c_{34}=10$, the achievable
region of TC is strictly larger than RC. When
$c_{12}=c_{34}=\infty$, both schemes meet their respective outer
bound. However, due to the single user half-duplex relay channel
capacity constraints (see (\ref{eq:single1})), RC achieves less
single user rates under the assumed channel conditions.

Bridging the gap between the outer bound and the achievable region for finite cooperative channel gains should be considered in future work.

\bibliographystyle{ieeetr}
\bibliography{cooperation}

\end{document}